\documentclass[12pt]{article}
\usepackage{amssymb}
\usepackage[dvips]{epsfig}

\newcommand{\be}{\begin{equation}}
\newcommand{\ee}{\end{equation}}
\def\bea{\begin{eqnarray}}
\def\eea{\end{eqnarray}}

\newcommand{\stuck}{St$\ddot{u}$ckelberg }
\newcommand{\bn}{\begin{eqnarray}}
\newcommand{\en}{\end{eqnarray}}

\newcommand{\oA}{{\overline{A}}}

\newcommand{\p}{\partial}

\newcommand{\mn}{\mu\nu}

\newcommand{\nn}{\nonumber}

\newcommand{\wmn}{W_{\mu\nu}}
\newcommand{\wmnu}{W^{\mu\nu}}
\newcommand{\no}{\noindent}

\newcommand{\tb}{\tilde{B}}
\newcommand{\ta}{\tilde{A}}
\newcommand{\tq}{\tilde{q}}
\newcommand{\tw}{\tilde{W}}

\newcommand{\s}{\,\,\,\,}
\def\bea{\begin{eqnarray}}
\def\eea{\end{eqnarray}}

\newcommand{\beq}{\begin{eqnarray}}
\newcommand{\eeq}{\end{eqnarray}}
\begin{document}

\title{\textbf{Mass generation for abelian spin-1 particles via  a symmetric tensor }}
\author{D. Dalmazi\footnote{dalmazi@feg.unesp.br} and E. L. Mendon\c ca\footnote{elias.fis@gmail.com} \\
\textit{{UNESP - Campus de Guaratinguet\'a - DFQ} }\\
\textit{{Avenida Doutor Ariberto Pereira da Cunha, 333} }\\
\textit{{CEP 12516-410 - Guaratinguet\'a - SP - Brazil.} }\\}
\date{\today}
\maketitle

\begin{abstract}
In the topologically massive BF  model (TMBF) the photon becomes
massive via coupling to an antisymmetric tensor, without breaking
the $U(1)$ gauge symmetry . There is no need of a Higgs field. The
TMBF model is dual to a first-order (in derivatives) formulation
of the Maxwell-Proca theory where the antisymmetric field plays
the role of an auxiliary field. Since the Maxwell-Proca theory
also admits a first-order version which makes use of an auxiliary
symmetric tensor, we investigate here a possible generalization of
the TMBF model where the photon acquires mass via coupling to a
symmetric tensor. We show that it is indeed possible to build up
dual models to the Maxwell-Proca theory where the $U(1)$ gauge
symmetry is manifest without Higgs field, but after a local field
redefinition the vector field eats up the trace of the symmetric
tensor and becomes massive. So the explicit $U(1)$ symmetry can be
removed unlike the TMBF model.
\end{abstract}

\newpage

\section{Introduction}

In the usual description of massive spin-1 particles via a
Maxwell-Proca (MP) action the gauge symmetry is explicitly broken.
It is of interest to search for alternatives to the Higgs
mechanism to preserve the gauge symmetry while generating a mass
for a spin-1 particle specially for the nonabelian case. Here we
address this question in the simpler case of the abelian $U(1)$
gauge symmetry. Dualization methods can help in investigating this
problem. It is convenient for those methods to rewrite the Maxwell
action in a first-order form by using auxiliary fields. In $D=1+1$
we can achieve that with help of a scalar field which interacts
with the vector field via a topological term $\phi \,
\epsilon^{\mu\nu}\p_{\mu}A_{\nu}$. By using the master action
approach of \cite{dj} as a dualization procedure it can be shown
\cite{an} that the first-order MP theory in $D=1+1$ is dual to a
local action with manifest $U(1)$ gauge symmetry. It corresponds
to the bosonized form of the Schwinger model whose effective
action, after elimination of the auxiliary scalar field, is
written down in our formula (\ref{schw}). Although non local, the
effective action is manifest $U(1)$ invariant.

In $D=2+1$ we replace the scalar field by a vector field $B_{\mu}$
and the topological coupling term becomes
$\epsilon^{\mu\nu\alpha}B_{\mu}\p_{\nu}A_{\alpha}$. After some
trivial field redefinition we end up, see master action in
\cite{jhep2} with equal masses, with a dual theory to  MP which
consists of a couple of noninteracting Maxwell-Chern-Simons
actions with the same mass but with opposite helicities. This
theory is manifest $U(1)$ symmetric and represents one massive
spin-1 particle with helicities $\pm 1$ just like the MP theory in
$D=2+1$.

In $D=3+1$ we can use an antisymmetric tensor with the so called
topological BF coupling
$\epsilon^{\mu\nu\alpha\beta}B_{\mu\nu}\p_{\alpha}A_{\beta}$. The
theory dual to MP is the topologically massive BF model (TMBF),
also named Cremmer-Scherk model \cite{cs}. It can be obtained from
the first-order MP theory via both master action \cite{botta} and
Noether gauge embedment \cite{menezes}. The TMBF model is unitary
\cite{abl} and explicitly $U(1)$ invariant. Unfortunately, as
shown in \cite{hlssvv}, a nonabelian generalization of the TMBF
model without extra fields will necessarily lead to power-counting
nonrenormalizable couplings as in \cite{ft}, see however,
\cite{lahiri,hls} where the extra field is nonpropagating  and
\cite{sav} for a recent suggestion which makes use of tensor gauge
fields. In \cite{bottaij} the geometrical origin of tensor gauge
connections is investigated. Thus, it is welcome to try
alternatives to the TMBF model. Here we follow this route for the
abelian $U(1)$ case as a laboratory for a possible non-abelian
generalization.

In fact, in \cite{kmu} there appears a new first-order form of the Maxwell action which makes use of a symmetric
auxiliary field $W_{\mu\nu}=W_{\nu\mu}$. By adding the Proca mass term we build up a first-order version of the
MP theory, see \cite{ds}. Now we have the coupling term $W_{\mu\nu}\p^{\mu}A^{\nu}$, though nontopological, this
term by itself has no particle content. In the next section we use this first-order formulation of the MP theory
as a starting point to obtain via master action and Noether gauge embedment alternative dual theories to the MP
theory. In section III we start with an Ansatz quadratic in the fields $A_{\mu}$ and $W_{\mu\nu}$ and
second-order in derivatives. We analyze its particle content and the presence of $U(1)$ gauge symmetry. In
section IV we draw our conclusions.

\section{Master action and Noether gauge embedment}

We begin with an alternative derivation of the TMBF model as given in \cite{botta}. We first define in $D=3+1$ a
master action depending on four different fields\footnote{In this work we use mostly plus $D$-dimensional
signature $\eta_{\mu\nu}=(-,+,\cdots , +)$} :

\bea S_M\lbrack A,\ta,B,\tb \rbrack = &-& \int \, d^4x
\left\lbrack \frac{m^2}2 A_{\mu}A^{\mu} + \frac{m^2}4
B_{\mu\nu}B^{\mu\nu} + \frac m2
\epsilon^{\mu\nu\alpha\beta}B_{\mu\nu}\p_{\alpha}A_{\beta} \right.
\nn
\\ &-& \left. \frac m2
\epsilon^{\mu\nu\alpha\beta}\left( B_{\mu\nu} -
\tb_{\mu\nu}\right) \p_{\alpha}\left( A_{\beta}-
\ta_{\beta}\right) \right\rbrack  \label{sm1}\eea

\no The first three terms of (\ref{sm1}) correspond to a
first-order version of the Proca theory. So their particle content
is one massive spin-1 particle. The last term of (\ref{sm1}) mixes
the fields $(A,B)$ with the dual ones $(\ta,\tb)$ . After the
shifts $\tb_{\mu\nu} \to \tb_{\mu\nu} + B_{\mu\nu}$ and $\ta_{\mu}
\to \ta_{\mu} + A_{\mu}$ it  decouples from the first three terms.
Its spectrum is empty (topological $BF$-term). So we conclude that
the spectrum of (\ref{sm1}) consists of one massive spin-1
particle. The mixing term does not contribute to the particle
content as usually in the master action approach.

On the other hand, if we Gaussian integrate the $(A,B)$ fields we obtain the TMBF model in terms of the dual
fields $(\ta,\tb)$:

\be S_{TMBF} = \int \, d^4x \left\lbrack - \frac 14
\tilde{F}_{\mu\nu}\tilde{F}^{\mu\nu}  + \frac
1{12}\tilde{H}_{\mu\nu\lambda}\tilde{H}^{\mu\nu\lambda} + \frac m4
\epsilon^{\mu\nu\alpha\beta}\tilde{B}_{\mu\nu}\tilde{F}_{\alpha\beta}
\right\rbrack \quad. \label{tmbf} \ee

\no Where $\tilde{F}_{\mu\nu}=\p_{\mu} \ta_{\nu} - \p_{\nu}
\ta_{\mu} $ and
$\tilde{H}_{\mu\nu\lambda}=\p_{\mu}\tb_{\nu\lambda} +
\p_{\nu}\tb_{\lambda\mu} + \p_{\lambda}\tb_{\mu\nu} $. The action
$S_{TMBF}$ is invariant under the independent gauge
transformations:

\bea \delta_{\phi} \ta_{\mu} = \p_{\mu} \phi \quad &;& \quad
\delta_{\phi} \tb_{\mu\nu} = 0 \label{gt1} \\
\delta_{\Lambda} \ta_{\mu} =0 \quad &;& \quad \delta_{\Lambda} \tb_{\mu\nu} = \p_{\mu} \Lambda_{\nu} - \p_{\nu}
\Lambda_{\mu} \quad . \label{gt2} \eea

\no It is convenient for future comparison to recall the equations
of motion of the TMBF model. Skipping the tildes for convenience
we have from (\ref{tmbf}):

\bea \p_{\alpha}H^{\mu\nu\alpha} + m\,
\epsilon^{\mu\nu\alpha\beta}\p_{\alpha}A_{\beta} &=& 0
\label{eqm1} \\
\p_{\nu}F^{\mu\nu} - \frac m6
\epsilon^{\mu\nu\alpha\beta}H_{\nu\alpha\beta} &=& 0 \label{eqm2}
\eea

\no Following \cite{abl}, we first solve (\ref{eqm1}). Recalling that in four dimensions and for vanishing mass
a two-form is dual to a scalar field ($\eta$) we have the general solution $H_{\alpha\beta\gamma} =
\epsilon_{\alpha\beta\gamma\delta}\left(\p^{\delta}\eta - m\, A^{\delta}\right) $. Back in (\ref{eqm2}) we have

\be \p_{\mu}F^{\mu\nu} - m^2 A^{\mu} + m\, \p^{\mu}\eta = 0
\label{procaeta} \ee

\no Since $\delta_{\phi} H_{\alpha\beta\gamma} = 0$ we must have
$\delta_{\phi} \eta = m\, \phi $. In terms of the gauge invariant
vector field $\overline{A}^{\mu} = A^{\mu} - \p^{\mu}\eta/m$ the
equation (\ref{procaeta}) becomes the well known Maxwell-Proca
equation without manifest $U(1)$ gauge invariance :

\be \p_{\mu}\overline{F}^{\mu\nu} - m^2 \oA^{\mu}  = 0 \quad ,
\label{proca}  \ee

\no In summary, the gauge invariant equations of motion (\ref{eqm1}) and (\ref{eqm2}) show that $B_{\mu\nu}$ can
be eliminated in terms of a gauge invariant vector field which satisfies the Proca equation and consequently we
have only one massive spin-1 particle in the spectrum of the TMBF model as expected from the master action
approach.

In the first three terms of (\ref{sm1}) we have written the
Maxwell-Proca theory in a first-order form with the help of a
totally antisymmetric rank two tensor ($B_{\mu\nu}$). It is
possible \cite{kmu,ds} to replace the two form field by a totally
symmetric tensor $W_{\mu\nu}=W_{\nu\mu}$ and a master action
similar to (\ref{sm1}) can be written down in arbitrary
$D$-dimensions as

\bea S_M\lbrack A,\ta,W,\tw \rbrack &=& \int\, d^D\, x \left\lbrack W^{\mn}W_{\mn} - \frac{W^2}{D-1} + 2 \,
W^{\mn}\p_{(\mu}A_{\nu )}  - \frac{m^2}2
A^{\mu}A_{\mu} \right. \nn\\
&-& \left. 2 \, \left(W^{\mn} - \tw^{\mu\nu}\right)\left(\p_{(\mu}A_{\nu )} - \p_{(\mu}\ta_{\nu )}\right)
\right\rbrack \label{sm2} \eea

\no The first four terms of (\ref{sm2}) correspond to a
first-order form of the Maxwell-Proca theory while the last term
mixes the $(A,W)$ fields with the duals $(\ta,\tw)$. After the
shift $\ta_{\mu} \to \ta_{\mu} + A_{\mu}$ and $ \tw_{\mu\nu} \to
\tw_{\mu\nu} + W_{\mu\nu}$ the last term of (\ref{sm2}) decouples
and becomes ${\cal L}_{\ta,\tw} = -2\, \tw^{\mn}\p_{(\mu}\ta_{\nu
)}$. Thus, the particle content of the master action (\ref{sm2})
corresponds to one massive spin-1 particle plus the content of
${\cal L}_{\ta,\tw}$. Minimizing the action\footnote{We quit the
tildes for while.} $S_{A,W} = \int d^D x \, {\cal L}_{A,W}$ we
have the equations of motion:

\be \p_{\mu} A_{\nu} + \p_{\nu} A_{\nu} = 0 \label{eqm3} \ee

\be  \p^{\mu}W_{\mu\nu} = 0 \label{eqm4} \ee

 \no It is easy to
convince oneself, assuming vanishing fields at infinity, that the
solution of (\ref{eqm3}) is trivial $A_{\mu}=0$ while (\ref{eqm4})
is solved \cite{dt} by $W_{\mu\nu} =
\p^{\alpha}\p^{\beta}R_{\mu\alpha\beta\nu}$ where
$R_{\mu\alpha\beta\nu}$ is a tensor with the index symmetries of
the Riemann curvature tensor but otherwise arbitrary. However,
since the action $S_{A,W}$ is itself invariant under
$\delta_{\Lambda}W_{\mu\nu} =
\p^{\alpha}\p^{\beta}\Lambda_{\mu\alpha\beta\nu}$ where
$\Lambda_{\mu\alpha\beta\nu}$ has the same properties of
$R_{\mu\alpha\beta\nu}$ we can say that the general solution of
(\ref{eqm4}) is pure gauge. Therefore, the last term of
(\ref{sm2}) has no particle content and the whole master action
(\ref{sm2}) contains only one massive spin-1 particle in the
spectrum. Following the master action approach if we Gaussian
integrate over the fields $(A,W)$ we have the dual action to the
first-order Maxwell-Proca model:

\be S^* = \int\, d^D\, x \left\lbrack - \frac 14 F_{\mu\nu}^2(\ta ) + \frac 2{m^2} \left(
\p^{\alpha}\tw_{\alpha}^{\nu}\right)^2 -2\,\tw^{\mn}\p_{(\mu}\ta_{\nu )} \right\rbrack  \label{sdual1} \ee

\no The action $S^*$ is invariant under the gauge transformations
$\delta_{\Lambda}\tw_{\mu\nu} =
\p^{\alpha}\p^{\beta}\Lambda_{\mu\alpha\beta\nu}$ with
$\delta_{\Lambda}\ta_{\mu} =0$. The equations of motion of
(\ref{sdual1}) are :

\bea \p_{\mu} \tq_{\nu} + \p_{\nu} \tq_{\mu} &=& 0 \label{eqm5} \\
\Box \theta_{\mu\nu} \ta^{\nu} - m^2 \ta_{\mu} + m^2 \tq_{\mu} &=&
0 \label{eqm6} \eea

\no where $\tq_{\mu} = \ta_{\mu} + 2
\p^{\alpha}\tw_{\alpha\mu}/m^2 $ is gauge invariant. As in
(\ref{eqm3}), due to the boundary conditions at infinity, we have
the solution $\tq_{\mu}=0 $ of (\ref{eqm5}) which allows us to
eliminate $\tw_{\mu\nu}$ in terms of $\ta_{\mu}$ up to gauge
transformations, i.e., $\p^{\alpha}\tw_{\alpha\beta} = - m^2
\ta_{\beta}/2 $. Back in (\ref{eqm6}) we recover the Maxwell-Proca
equation confirming that $S^*$ contains only one massive particle
of spin-1 in the spectrum. Although $S^*$ is equivalent to the
Maxwell-Proca theory,  the dual model (\ref{sdual1}) has no $U(1)$
gauge symmetry (differently from $S_{TMBF}$). The key point is
that the mixing term ($BF$ term) in the master action (\ref{sm1})
carries $U(1)$ gauge symmetry contrary to  the mixing term of the
master action (\ref{sm2}). We conclude that a natural application
of the master action approach to the first-order Maxwell-Proca
theory obtained with help of a symmetric tensor $W_{\mu\nu}$ does
not lead us to a theory with explicit $U(1)$ gauge invariance.

Another way of obtaining $S_{TMBF}$  from the Maxwell-Proca theory
is by means of a Lagrangian Noether gauge embedding (NGE)
procedure as in \cite{menezes,ilha}, see also \cite{hs} for a
Hamiltonian embedding. Let us repeat the same Lagrangian procedure
here. The first four terms of (\ref{sm2}) define the first-order
Maxwell-Proca theory:

\be S^{(0)} = \int\, d^D\, x \left\lbrack W^{\mn}W_{\mn} -
\frac{W^2}{D-1}  + 2 \, W^{\mn}\p_{(\mu }A_{\nu )} - \frac{m^2}2
A^{\mu}A_{\mu}\right\rbrack \label{s0} \ee

\no The first three terms of (\ref{s0}) are invariant under the
$U(1)$ gauge transformations:

\be \delta_{\phi} A_{\mu} =  \p_{\mu} \phi  \quad ; \quad \delta_{\phi}W_{\mu\nu}  =  \Box \theta_{\mu\nu}\phi
\quad . \label{gtphi}\ee

\no  Where we define the projection operators:

\be \theta_{\alpha\beta} = \left( \eta_{\alpha\beta} -
\omega_{\alpha\beta} \right) \quad , \quad \omega_{\alpha\beta} =
\frac{\p_{\alpha}\p_{\beta}}{\Box} \label{theta} \ee

\no In order to preserve the $U(1)$ symmetry we can modify the
action $S^{(0)}$ by adding  linear terms in auxiliary fields:

\be S^{(1)} = S^{(0)} + \int\, d^D\, x \left( K_{\mu}B^{\mu} +
M_{\mu\nu}C^{\mu\nu} \right)  \quad . \label{s1} \ee

\no where the Euler tensors are given by

\bea K_{\mu} &=& \frac{\delta S^{(0)}}{\delta A^{\mu}} = -
m^2A_{\mu} - 2 \p^{\nu}W_{\mu\nu} \label{euler1} \\
M_{\mu\nu} &=& \frac{\delta S^{(0)}}{\delta W^{\mu\nu}} = 2
\left\lbrack \p_{(\mu}A_{\nu )} + W_{\mu\nu} - \eta_{\mu\nu} \frac
W{D-1} \right\rbrack \label{euler2} \eea

\no The auxiliary fields must transform with the opposite sign of
the original fields such that their variations in (\ref{s1})
compensate the variation of $S^{(0)}$,

\be \delta_{\phi} B^{\mu} = - \p^{\mu} \phi  \quad ; \quad \delta_{\phi}C^{\mu\nu}  = - \Box \theta^{\mu\nu}\phi
\quad . \label{gtaux}\ee

\no Since $\delta_{\phi}M_{\mu\nu} =0 $ and $\delta_{\phi} K_{\mu}
= - m^2\p_{\mu}\phi = m^2 \delta_{\phi} B_{\mu} $ we have

\be \delta_{\phi}S^{(1)} =  \int\, d^D\, x \delta_{\phi}\left( m^2 \frac{B^{\mu}B_{\mu}}2 \right) \ee

\no So we deduce the gauge invariant action:

\be S^{(2)} = S^{(0)}  + \int\, d^D\, x \left( K_{\mu}B^{\mu} -
m^2 \frac{B^{\mu}B_{\mu}}2  + C^{\mu\nu}M_{\mu\nu} \right)
\label{s2} \ee

\no After a functional integral over the auxiliary fields we get a functional delta function $\delta
(M_{\mu\nu})$ which allows us to further integrate over $W_{\mu\nu}$. We end up with the following gauge
invariant effective action for the vector field

\be S_{P} = \frac 14 \int\, d^D\, x F_{\mu\nu}\left(1 - \frac{\Box}{m^2} \right) F^{\mu\nu}  \label{pod} \ee

\no The action (\ref{pod}) is the Podolsky \cite{pod} action up to an overall sign. Its equations of motion can
be written as

\be \Box \left( \Box - m^2 \right)A_{\mu}^T =0 \label{eqmpod} \ee

\no where   $A_{\mu}^T = \theta_{\mu\nu}A^{\nu} $ satisfies
$\p^{\mu}A_{\mu}^T = 0 $ which altogether with $\left( \Box - m^2
\right)A_{\mu}^T =0 $ are equivalent to the Maxwell-Proca
equations as expected from the embedment procedure. However,
(\ref{eqmpod}) also contain massless solutions to the usual free
Maxwell equations $\Box \theta_{\mu\nu}A^{\nu} = 0 $. An analysis
of the sign of the imaginary part of the residues of both massless
and massive poles of the propagator coming from (\ref{pod})
reveals that we have one massive physical particle and one
massless ghost which violates unitarity. This is not completely
unexpected from the point of view of the NGE procedure as
explained in \cite{baeta}.

In summary the master action and the NGE methods have led us to
different results and none of them is satisfactory. In the next
section we use another approach for a broader investigation of
this question.

\section{ A general Ansatz}

Another way of figuring out the particle content of the TMBF model
is to integrate in the two form field $B_{\mu\nu}$  in the path
integral and obtain an effective action for the vector field
$A_{\mu}$. One ends up, see \cite{barcelos}, with a four
dimensional version of the well known Schwinger model which
appears in $D=1+1$ dimensions due to the non-conservation of the
axial current, namely :

\be \exp^{i S_{eff}\lbrack A \rbrack} = \int {\cal D}B_{\mu\nu}
\exp^{i S_{TMBF}\lbrack A,B \rbrack } \label{seff1} \ee

\no Where

\be S_{eff}\lbrack A \rbrack = S_{Schw} = - \frac 14 \int d^4 x \, F_{\mu\nu}\frac{\left(\Box - m^2
\right)}{\Box} F^{\mu\nu} \label{schw} \ee

\no The Schwinger model is of course $U(1)$ gauge invariant and a
careful analysis of the analytic properties of the propagator
reveals that we have only one massive (spin-1) particle in the
spectrum as in the initial TMBF model. In what follows we start
with a more general (second-order in derivatives) Ansatz for a
local quadratic action containing the fields
$(A_{\mu},W_{\mu\nu})$ and integrate over $W_{\mu\nu}$ in order to
deduce a $D$ dimensional effective action for the vector field.

Let us start with the Ansatz:

\bea S \left\lbrack A, W \right \rbrack &=& \int d^D x\left\lbrack
a \left( \p \cdot A \right)^2 + b \left( \p_{(\mu}A_{\nu
)}\right)^2 + c_1 \left( \p^{\nu}W_{\mu\nu} \right)^2 + c_2\,
\p^{\nu}W \p^{\mu}\wmn \right. \nn\\
&+& \left. c_3 \, \p^{\mu}W \p_{\mu}W + c_4 \, \p^{\alpha} \wmn
\p_{\alpha}\wmnu + d \, \wmn \wmnu + e \, W^2 \right. \nn \\&+&
\left. f\, \wmn \p^{\mu}A^{\nu} + g \, W \p \cdot A \right\rbrack
\label{ansatz} \eea

\no where $(a,b,c_i,e,f,g)$ are so far unknown real constants. We
can rewrite the Ansatz as:

\be S  \left\lbrack A, W \right \rbrack = \int d^D x \left\lbrack
a \left( \p \cdot A \right)^2 + b \left( \p_{(\mu}A_{\nu
)}\right)^2 + \wmn G^{\mu\nu}_{\quad \alpha\beta}W^{\alpha\beta} +
W_{\alpha\beta} T^{\alpha\beta} \right\rbrack \quad.
\label{ansatz2} \ee

\no Where

\be T^{\alpha\beta} = f\,  \p^{(\alpha}A^{\beta )} + g\,
\eta^{\alpha\beta} \, \p \cdot A  \label{t} \ee

\bea G^{\mu\nu}_{\quad \alpha\beta} &=& \left\lbrace\left( d-c_4 \Box \right)P_{SS}^{(2)} + \left( d -
\frac{c_1\Box}{2} - c_4
\Box\right) P_{SS}^{(1)}   \right. \nn\\
&+& \left.  \left\lbrack d + e - (c_1 + c_2 + c_3 + c_4)\right\rbrack P_{WW}^{(0)} + \left\lbrack d - c_4\Box +
(e-c_3\Box)(D-1)\right\rbrack P_{SS}^{(0)} \right. \nn \\
&+& \left. \sqrt{D-1}\left(e - c_3\Box - \frac{c_2\Box}2 \right) \left(T_{SW}^{(0)} + T_{WS}^{(0)}
\right)\right\rbrace^{\mu\nu}_{\alpha\beta}  \label{propa} \eea

\no where the projection operators $P_{IJ}^{(s)}$ of spin-$s$  and the transition operators $T_{SW}^{(0)} \, ,
\, T_{WS}^{(0)}$ are defined as :

\be \left( P_{SS}^{(2)} \right)^{\lambda\mu}_{\s\s\alpha\beta} = \frac 12 \left(
\theta_{\s\alpha}^{\lambda}\theta^{\mu}_{\s\beta} + \theta_{\s\alpha}^{\mu}\theta^{\lambda}_{\s\beta} \right) -
\frac{\theta^{\lambda\mu} \theta_{\alpha\beta}}{D-1}  \quad , \label{ps2} \ee

\be \left( P_{SS}^{(1)} \right)^{\lambda\mu}_{\s\s\alpha\beta} = \frac 12 \left(
\theta_{\s\alpha}^{\lambda}\,\omega^{\mu}_{\s\beta} + \theta_{\s\alpha}^{\mu}\,\omega^{\lambda}_{\s\beta} +
\theta_{\s\beta}^{\lambda}\,\omega^{\mu}_{\s\alpha} + \theta_{\s\beta}^{\mu}\,\omega^{\lambda}_{\s\alpha}
 \right) \quad , \label{ps1} \ee

\be \left( P_{SS}^{(0)} \right)^{\lambda\mu}_{\s\s\alpha\beta} = \frac 1{D-1} \,
\theta^{\lambda\mu}\theta_{\alpha\beta} \quad , \quad \left( P_{WW}^{(0)} \right)^{\lambda\mu}_{\s\s\alpha\beta}
= \omega^{\lambda\mu}\omega_{\alpha\beta} \quad , \label{psspww} \ee

\be \left( T_{SW}^{(0)} \right)^{\lambda\mu}_{\s\s\alpha\beta} =
\frac 1{\sqrt{D-1}}\, \theta^{\lambda\mu}\omega_{\alpha\beta}
\quad , \quad  \left( T_{WS}^{(0)}
\right)^{\lambda\mu}_{\s\s\alpha\beta} = \frac 1{\sqrt{D-1}}\,
\omega^{\lambda\mu}\theta_{\alpha\beta} \quad , \label{pswpws} \ee

\no From (\ref{ansatz2}), integrating over the fields $\wmn $ in
the path integral we obtain the effective action

\be S_{eff} \left\lbrack A \right\rbrack = \int d^D x \left\lbrack
a \left( \p \cdot A \right)^2 + b \left( \p_{(\mu}A_{\nu
)}\right)^2 - \frac 14 T_{\mu\alpha}(A) \left( G^{-1}
\right)^{\mu\alpha}_{\,\,\gamma\beta}T^{\gamma\beta}(A) \,
,\right\rbrack \label{sef2} \ee

\no where, suppressing the indices for convenience, we have

\bea G^{-1}  &=& \frac{P_{SS}^{(2)}}{(d-c_4\Box)} +
\frac{P_{SS}^{(1)}}{d-\Box\left(c_4+\frac{c_1}2\right) } +
\frac{\left\lbrack d - c_4\Box +
(e-c_3\Box)(D-1)\right\rbrack P_{WW}^{(0)}}K  \nn \\
&+& \frac{\left\lbrack d + e - (c_1 + c_2 + c_3 + c_4)\Box
\right\rbrack P_{SS}^{(0)}}K  + \frac{\sqrt{D-1}}K (e - c_3\Box -
\frac{c_2\Box}2 )(T_{SW}^{(0)} + T_{WS}^{(0)}) \label{gmenos1}
\eea

\no With

\bea K &=& \left\lbrack d + e - (c_1 + c_2 + c_3 +
c_4)\Box\right\rbrack \left\lbrack d - c_4\Box +
(e-c_3\Box)(D-1)\right\rbrack \nn \\
&-&  (D-1)\left(e - c_3\Box - \frac{c_2\Box}2 \right)^2 \,
.\label{K} \eea

\no Working out the expression (\ref{sef2}) we have

\be S_{eff} \left\lbrack A \right\rbrack = \int d^D x  \left\lbrack a \left( \p \cdot A \right)^2 + b \left(
\p_{(\mu}A_{\nu )}\right)^2 + (\p \cdot A)  H(\Box ) (\p \cdot A) -   \frac{1}{16} F_{\mu\nu}\frac {f^2}{d-
\Box\left(c_4+\frac{c_1}2\right)}F^{\mu\nu}\right\rbrack \\ \label{sef3} \ee

\no where

\bea H(\Box ) &=& \frac{-1}{4\,K}\left\lbrace (D -
1)g^2\left\lbrack d + e - (c_1 + c_2 + c_3 + c_4)\Box
\right\rbrack + (f + g)^2\left\lbrack d - c_4 \Box + (e - c_3 \Box
)(D - 1)\right\rbrack \right. \nn\\ &+& \left. 2(D - 1)g(f +
g)\left\lbrack (c_3 + c_2/2)\Box - e\right\rbrack \right\rbrace
\label{H} \eea

\no In order to have $U(1)$ gauge invariance in (\ref{sef3}) the constants in our Ansatz (\ref{ansatz}) must be
such that

\be H(\Box ) = - (a + b) \label{hab} \ee

\no Consequently we end up with the gauge invariant theory

\be S_{eff} \left\lbrack A \right\rbrack = - \frac 1{16} \int d^D
x F_{\mu\nu}\frac{4\, b \left\lbrack \left(c_4 +
\frac{c_1}2\right)\Box - d \right\rbrack + f^2}{d- \left(c_4 +
\frac{c_1}2 \right)\Box} F^{\mu\nu} \quad . \label{seffinal} \ee

\no By adding a gauge fixing term we can obtain the propagator and calculate the saturated two point amplitude
in momentum space $A(k)$ from which we can read off the particle content of the theory. Explicitly,

\bea A(k) &=& J^*_{\mu}(k) \left\langle A^{\mu}(-k) A^{\nu}(k)
\right\rangle J_{\nu}(k) = - \frac i2 J^*_{\mu}(k) \left\lbrack
G^{-1}(k) \right\rbrack^{\mu\nu}  J_{\nu}(k) \nn\\ &=& -
 \frac i2 \frac{J^* (k) \cdot J(k) \left\lbrack \left( c_4 + \frac{c_1}2 \right)k^2 + d \right\rbrack }{k^2
 \left\lbrack 4\, b \left( c_4 + \frac{c_1}2 \right)k^2 + 4 \, b \, d - f^2 \right\rbrack } \label{ak} \eea

\no Note that the contribution of the gauge fixing term $\lambda
\left( \p \cdot A \right)^2 $ drops out from $A(k)$ due to the
transverse nature of the sources ($k \cdot J =0$) as required by
gauge invariance.

 We may have one or two poles in $A(k)$. Since our aim is to obtain only one physical massive  particle in
 the spectrum we impose henceforth :

\be d=0 \quad  ; \quad  b \, \left( c_4 + \frac{c_1}2 \right) \ne
0 \quad ; \quad f\ne 0  \label{cond1} \ee

\no In this case:

\be A(k) = - \frac{i\, J^* \cdot J }{8\, b \left( k^2 + m^2
\right) } \quad . \label{ak2} \ee

\no Where

\be m^2 = - \frac{f^2}{4\, b \left( c_4 + c_1/2 \right)} \quad . \label{mass} \ee

 \no The imaginary part of the
residue of $A(k)$ at the pole $k^2=-m^2$ becomes $-J^*(k) \cdot
J(k)/(8\, b)$ evaluated at $k^2=-m^2$. In the rest frame $k_{\mu}
= (m,0,\cdots,0)$, due to $k \cdot J (k) =0$, we must have $J_0(k)
=0$. So we can easily check that the frame  independent quantity
$J^*(k) \cdot J(k)$ is positive. Consequently,  in order to have a
physical particle as required by unitarity ($ImRes(A(k))> 0 $) and
be free of tachyons, see ({\ref{mass}), we must further assume
that :

\be b < 0 \quad ; \quad   c_4 + \frac{c_1}2 > 0 \label{cond2} \ee

\no According to the above requirements the effective action
$(\ref{seffinal})$ becomes exactly, fixing $b=-1$, the Schwinger
model effective action (\ref{schw}). Clearly, we have to inspect
the restrictions imposed by the gauge invariance condition
(\ref{hab}). Namely,

\be (D-1)\, e = 0 \label{gic1} \ee \be \left(a + b \right)
\left\lbrack (D-1)\left( c_2 f - 2 c_1 g \right) - 2 c_4 \left( f
+ D\, g \right ) \right\rbrack  = 0 \label{gic2} \ee \be
(D-1)f^2c_3 = g(D-1)\left( c_2 f - c_1 g \right) - c_4\left( f^2 +
2\, f \, g + D\, g^2 \right) \quad . \label{gic3} \ee

\no  For future use we recall that in deducing
(\ref{gic1}),(\ref{gic2}) and (\ref{gic3}) we have assumed $d=0$,
$c_4 \ne 0$, $c_4 + c_1/2 \ne 0$ and that $K\ne 0$ which means,
using $e=0$ according to (\ref{gic1}), that

\be K = \Box^2 \left\lbrace \left(c_1 + c_2 + c_3 + c_4 \right)\left\lbrack (D-1)c_3 + c_4\right\rbrack -
(D-1)\left( c_3 + \frac{c_2}2\right)^2 \right \rbrace  \ne 0 \label{K2} \ee

Although there are several solutions for (\ref{gic2}) and
(\ref{gic3}) some of them are related via trivial field
redefinitions in our Ansatz (\ref{ansatz}). Here we stick to the
simplest case. First of all we solve (\ref{gic2}) by choosing
$b=-a$ which implies that the first two terms of (\ref{ansatz})
build up the Maxwell Lagrangian density. So the gauge symmetry is
present before the integration over the $W_{\mu\nu}$ fields. This
is also the case of the TMBF model. For definiteness we choose
$a=1$ and $b=-1$. Moreover, if $b=-a$  the field redefinition
$A_{\mu} \to A_{\mu} + c_2 \p_{\mu}W/f $ in the Ansatz
(\ref{ansatz}) will bring $c_2 \to 0$. The coefficient $c_3$ also
changes but we rename it $c_3$ again without loss of generality.
Regarding $c_4$, we might think of choosing $c_4=0$ for
simplicity. However, since $d=0$, that leads to a zero in the
denominator of (\ref{gmenos1}). This indicates the appearance of a
gauge symmetry. Indeed, if $c_4=0$, the Ansatz (\ref{ansatz})
becomes invariant under any transformation which preserves both
the trace $W$ and $\p^{\mu}W_{\mu\nu}$. They are given by the
local transformations:

\be \delta_{\Lambda} W_{\mu\nu} = \left\lbrack \Box^2
P_{SS}^{(2)}\right\rbrack_{\mu\nu}^{\quad \alpha\beta}
\Lambda_{\alpha\beta} \quad \to \delta_{\Lambda} W = 0
\label{spin2gt} \ee

\no where $\Lambda_{\alpha\beta}=\Lambda_{\beta\alpha}$ is an
arbitrary symmetric tensor. In practice we can set $c_4=0$ but in
order to integrate over $W_{\mu\nu}$ the term $c_4\,
\p^{\mu}W_{\alpha\beta} \p_{\mu}W^{\alpha\beta} $ must be replaced
by a local gauge fixing term like

\be {\cal L}_{GF}^{(2)} = \lambda_2 \left( \Box^2 \left\lbrack
P_{SS}^{(2)}\right\rbrack_{\mu\nu}^{\quad \alpha\beta}
W_{\alpha\beta} \right)^2 \label{gf2} \ee

\no Now the denominator $(d-c_4\Box)$ of the first term of
(\ref{gmenos1}) is replaced by $\lambda_2 \Box^4$. Since
$T_{\mu\nu}\left\lbrack P_{SS}^{(2)}
\right\rbrack^{\mu\nu}_{\,\alpha\beta}T^{\alpha\beta} =0$ the
effective action, see (\ref{sef2}), does not depend on the
arbitrary real constant $\lambda_2$ as expected. Returning to our
$U(1)$ gauge invariance conditions, from (\ref{gic3}) we must have
$ c_3 = -(g/f)^2 c_1$. Thus, we may choose

\be a=1 \quad ; \quad b=-a=-1 \quad ; \quad c_2=0=c_4 \quad ;
\quad c_3 = -\left( \frac gf \right)^2 c_1  \quad , \label{cond3}
\ee

\no which leads to the model

\be S_I  = \int d^D x \left\lbrace -\frac 14 F_{\mu\nu}^2 + f\,
W_{\mu\nu} \p^{\mu}A^{\nu} + g\, W \, \p \cdot A + c_1
\left\lbrack \left( \p^{\mu}W_{\mu\nu}\right)^2 - \frac{g^2}{f^2}
\p^{\mu}W\p_{\mu}W \right\rbrack \right\rbrace \label{sg} \ee

\no We must have, see (\ref{cond1}),(\ref{cond2}) and
(\ref{cond3}), $c_1 >0$ and $f \ne 0$ while $g$ is an arbitrary
real constant. Unfortunately, the action $S_I$ is not invariant
under usual $U(1)$ gauge transformation in general but rather
under a higher derivative form of it,

\bea \delta_{\phi}^{hd}A_{\mu} &=& \p_{\mu}\Box\phi \label{gshda} \\
\delta^{hd}_{\phi}W_{\mu\nu} &=& \frac{f}{2\,
c_1(D-1)g}\left\lbrack \left(f+g\, D\right)\p_{\mu}\p_{\nu}\phi -
\left(f + g \right) \eta_{\mu\nu} \Box\phi \right\rbrack
\label{gshdw} \eea

\no Furthermore, if $f \ne - g \, D$ we can redefine the fields
according to

\be A_{\mu} = \ta_{\mu} -  \frac{2\, c_1 g}{f(f+g\, D)}\p_{\mu}
\tilde{W} \label{fr1} \ee

\be W_{\mu\nu} = \tilde{W}_{\mu\nu} - \frac g{f+ g\, D}
\eta_{\mu\nu} \tilde{W} \label{fr2} \ee \no which is equivalent to
set $g=0$ in $S_I$.  The new vector field $\ta_{\mu}$ is gauge
invariant  ($ \delta_{\phi}^{hd}\ta_{\mu} =0$) while
$\delta_{\phi}^{hd}\tilde{W}_{\mu\nu} = f(f + g\, D)\Box
\theta_{\mu\nu}\phi/\left\lbrack 2\, c_1(1-D)g\right\rbrack $.
This transformation preserves $\p^{\mu}\tw_{\mu\nu}$.

The $U(1)$ gauge symmetry is no longer manifest in $S_I(g=0)$.
However, in the $W$ sector we have a larger symmetry now since the
trace $W$ is absent and any transformation which preserves
$\p^{\mu}\tw_{\mu\nu}$ is a symmetry. So the $U(1)$ symmetry moves
to the $W$ sector. The action $S_I(g=0)$ is equivalent, with the
normalization $c_1=2/m^2$ and $f=-2$, to the dual model
(\ref{sdual1}) obtained from the MP theory via master action. It
is surprising to end up without explicit $U(1)$ gauge symmetry
after imposing the gauge invariance conditions
(\ref{gic1}),(\ref{gic2}) and (\ref{gic3}). However, the
derivation of those conditions requires $K\ne 0$ which is not
valid for $g=0$ since, see (\ref{cond3}), in this case
$c_2=0=c_4=c_3$.

On the other hand, if  $f = -D\, g$ we can choose, recalling
(\ref{mass}),  without loss of generality
 $f = m^2 = - D\, g $ and $c_1= m^2/2$. Back in (\ref{sg}) we have:

\be S_{II} = \int d^D x \left\lbrack -\frac 14 F_{\mu\nu}^2 + m^2
\left( W_{\mu\nu} - \frac WD \eta_{\mu\nu} \right) \p^{\mu}A^{\nu}
+  \frac{m^2}2 \left(
\p^{\mu}W_{\mu\nu}\p^{\alpha}W_{\alpha}^{\,\, \nu} - \frac{1}{D^2}
\p^{\mu}W\p_{\mu}W\right) \right\rbrack \label{sdual2} \ee

\no The action $S_{II}$ is explicitly invariant under usual
(first-order) $U(1)$ gauge transformations:

\be \delta_{\phi} A_{\mu} = \p_{\mu} \phi \quad ; \quad \delta_{\phi} W_{\mu\nu} = \eta_{\mu\nu} \phi
\label{u1} \ee

\no After adding an appropriate gauge fixing term like (\ref{gf2})
and integrating over $W_{\mu\nu}$ in the path integral we end up
with the effective action of the Schwinger type, see (\ref{schw}).
Thus, $S_{II}$ is a new action dual to the Maxwell-Proca theory
with manifest usual $U(1)$ symmetry. It is important to notice
however, that since $\delta_{\phi}W=D \phi $ we can always change
variables to a gauge invariant vector field $A_{\mu} \to
A_{\mu}-\p_{\mu}W/D$ and loose the manifest $U(1)$ symmetry. The
action now becomes:

\be S_{II-b} = \int d^D x \left\lbrack -\frac 14 F_{\mu\nu}^2 +
m^2 \left( W_{\mu\nu} - \frac WD \eta_{\mu\nu} \right)
\p^{\mu}A^{\nu} +  \frac{m^2}2 \left( \p^{\mu}W_{\mu\nu} -
\frac{\p_{\nu}W}{D} \right)^2 \right\rbrack \label{sdual2b} \ee

\no One can say that the initial massless vector field $A_{\mu}$
has eaten up the trace $W$ and became massive as in the usual
Stuckelberg mechanism. Notice also that in $S_{II-b}$ only the
traceless piece of $W_{\mu\nu}$ effectively appears contrary to
(\ref{sdual2}). The action $S_{II-b}$ is invariant under the
spin-2 local transformations (\ref{spin2gt}) but since the whole
$U(1)$ symmetry is shifted to the $W_{\mu\nu}$ sector. The
quadratic terms in $W_{\mu\nu}$ are also invariant under Weyl
transformations $\delta_{\phi} W_{\mu\nu} = \eta_{\mu\nu} \phi $
which require another gauge fixing term. We can choose for
instance:

\be {\cal L}_{GF}^{(0)} = \lambda_0 \left\lbrack\Box^2
\left(P_{WW}^{(0)}\right)_{\alpha\beta}^{\,\,
\mu\nu}W_{\mu\nu}\right\rbrack^2 \label{gf0} \ee

After adding ${\cal L}_{GF}^{(2)}$ and ${\cal L}_{GF}^{(0)}$ to
(\ref{sdual2b}) we can integrate over $W_{\mu\nu}$ and obtain an
effective action, independent of both $\lambda_0$ and $\lambda_2$,
which becomes exactly the Maxwell-Proca theory:

\be {\cal L}_{eff}\left\lbrack A \right\rbrack = {\cal L}_{MP} =
-\frac 14 F_{\mu\nu}^2 - \frac{m^2}2 A_{\mu}A^{\mu} \quad .
\label{mp} \ee

\no It is easy to check that $K=0$ for (\ref{sdual2b}) which
explains the loss of manifest $U(1)$ gauge symmetry once again.
The equations of motion of (\ref{sdual2b}) can be written as

\be \Box \theta_{\mu\nu}A^{\nu} - m^2  v_{\mu} = 0 \quad ; \quad
v_{\mu} = \p^{\alpha}W_{\alpha\mu} - \frac{\p_{\mu}W}D
\label{dualeq1} \ee

\be  \p_{\mu} q_{\nu} + \p_{\nu} q_{\mu}  = \frac 2D \eta_{\mu\nu}
\p \cdot q \quad ; \quad q_{\mu} =  A_{\mu} - v_{\mu}
\label{dualeq2} \ee

\no Note that the vectors $A_{\mu}$, $v_{\mu}$ and consequently
$q_{\mu}$ are $U(1)$ gauge invariant. General coordinate
transformations in a flat space time changes the metric tensor
according to $\delta_{\xi} g_{\mu\nu} = \p_{\mu} \xi_{\nu} +
\p_{\nu} \xi_{\mu} $. Conformal transformations require that
$\delta_{\xi} g_{\mu\nu} = \Lambda \, g_{\mu\nu} $ whose trace
implies $\Lambda = 2(\p \cdot \xi )/D $. Therefore, the general
solution to the first equation of (\ref{dualeq2}) corresponds
exactly to conformal transformations

\be q_{\mu} = A_{\mu} - v_{\mu} = a_{\mu} + \Lambda_{\mu\nu}
x^{\nu} + \lambda\, x_{\mu} + 2 x_{\mu} \left( x \cdot c \right )
- x^2 c_{\mu} \quad , \label{ct} \ee

\no where the antisymmetric matrix $\Lambda_{\mu\nu}  $ and
$a_{\mu},b_{\mu}, c_{\mu}, \lambda $ are constant parameters.
Since the fields must vanish at infinity, all those constant
parameters must vanish. So $q_{\mu}= v_{\mu} - A_{\mu} = 0 $
allows us to replace $v_{\mu}$ by $A_{\mu}$ in (\ref{dualeq1})
which becomes, as expected from the effective action, the
Maxwell-Proca equation $ \Box \theta_{\mu\nu} A^{\nu} - m^2
A_{\mu} = 0 $.

After eliminating $v_{\mu}$ in terms of $A_{\mu}$ we are still
left with degrees of freedom in $W_{\mu\nu}$ which do not
contribute do the combination $v_{\mu}$ however, those are exactly
the pure gauge degrees of freedom related to the symmetries of
(\ref{sdual2b}). So the duality between (\ref{sdual2b}) and the
Maxwell-Proca theory is also established at classical level as
expected.

At this point one might ask whether it is possible to define a
unitary theory containing only one massive spin-1 particle
starting with quadratic terms in $W_{\mu\nu}$ and the vector field
$A_{\mu}$ such that the manifest $U(1)$ symmetry  can not be
removed by any local field redefinition as in the TMBF model. In
the unitary dual models $S_I$ and $S_{II}$ it was possible to
define a gauge invariant linear combination of the vector fields
$A_{\mu}$ and $\p_{\mu}W$ and remove the manifest $U(1)$ symmetry.

One might  blame the choice (\ref{cond3}) for the existence of a
$U(1)$ gauge invariant vector field which leads to the lack of
manifest gauge invariance in general. Next we give a symmetry
argument to show that even for the general Ansatz this will be
always possible. So let us return to the general Ansatz
(\ref{ansatz}) and address this question from the point of view of
gauge transformations. Namely, the $U(1)$ gauge transformation
which leave the Ansatz (\ref{ansatz}) invariant must be of the
general form

\be \delta_{\phi}A_{\mu} = \p_{\mu}\phi \quad ; \quad
\delta_{\phi}W_{\mu\nu} = r\, \phi \, \eta_{\mu\nu} + s \, \Box
\phi \, \eta_{\mu\nu} + t\, \p_{\mu}\p_{\nu}\phi \label{ggt} \ee

\no where $(r,s,t)$ are real constants. The variation of the
Ansatz includes the following independent terms:

\be \delta S  = \int d^Dx \left\lbrack 2\, r (D \, e\, + d ) W\,
\phi + r (f+D\, g)\p^{\mu}A_{\mu} \phi + \left(f-2\, r\, c_1 - D\,
r\, c_2 + 2\, d \, t\right)\p^{\mu}\p^{\nu}W_{\mu\nu} \phi +
\cdots \right\rbrack \label{deltas} \ee

\no Therefore, among other constraints, we have the following ones

\bea r (D \, e\, + d ) &=& 0 \label{c1} \\
r\left( f + D\, g \right) &=& 0 \label{c2} \\
 r\left( 2\, c_1 + D\, c_2 \right) -2\, d \, t  &=& f \label{c3} \eea

\no For only one massive particle in the spectrum we must have
$d=0$ and $f \ne 0$, therefore $r \ne 0$ so we can rescale $r \to
1$. It also follows that $e=0$ and $f=-D\, g$ which is in
agreement with our previous results $S_I$ and $S_{II}$ since we
have demanded usual (first-order) $U(1)$ transformations for the
vector field.

 On the other hand, the field redefinition

\be W_{\mu\nu} = \tw_{\mu\nu} + s\, \eta_{\mu\nu} \p \cdot A + t\,  \p_{(\mu}A_{\nu )} \label{fr} \ee

\no will absorb the $t$ and $s$ factors such that
$\delta_{\phi}W_{\mu\nu} = \eta_{\mu\nu} \phi $, i.e., we can set
$s=0=t$ in (\ref{ggt}). Therefore, we conclude that we are always
able to make a field redefinition  $ A_{\mu} = \ta_{\mu} +
\p_{\mu}W/D $ to  a gauge invariant vector field
$\delta_{\phi}\ta_{\mu}=0$ which jeopardizes the manifest $U(1)$
symmetry.

In fact, it is easy to show that the $U(1)$ symmetry will be
indeed lost after the field redefinition since there will be no
more contribution coming from $f
W_{\mu\nu}\p^{\mu}\delta_{\phi}\ta^{\nu}$ to the last (explicit)
term of (\ref{deltas}). Consequently, the new quadratic terms in
$W_{\mu\nu}$ must satisfy $c_2=-2\, c_1/D $. So the gauge
variation of the quadratic terms in $W_{\mu\nu}$ cancel out with
no  need of any contribution coming from the vector field. Since a
nonvanishing mass requires $f\ne 0$, it is clear that a possible
$U(1)$ variation $f W_{\mu\nu}\p^{\mu}\delta_{\phi}\ta^{\nu}= f
W_{\mu\nu}\p^{\mu}\p^{\nu}\phi $ can not be compensated by the
variation of $W_{\mu\nu}$ fields and the manifest $U(1)$ symmetry
is lost.

In practice we have checked that even for other choices different from (\ref{cond3}), it is always possible to
redefine the fields and end up without manifest $U(1)$ symmetry.

\section{Conclusion}

In the topologically massive BF model (TMBF), also named
Cremmer-Scherk model, the photon acquires mass without need of a
Higgs field while keeping the $U(1)$ gauge symmetry manifest in
the action. It is not possible in this case to remove the $U(1)$
symmetry from the action by any local field redefinition. In this
model the vector field is coupled to an antisymmetric tensor.
Motivated by the TMBF model we have investigated here the
possibility of generating mass for the photon, in a $U(1)$
invariant way,  by coupling the vector field to a symmetric rank-2
tensor instead. Since the TMBF model can be interpreted as a dual
version of a first-order formulation of the Maxwell-Proca theory,
we have applied standard dualization methods to a first-order form
of the Maxell-Proca theory which makes use of a symmetric tensor,
see \cite{kmu}. In particular, we have used the master action and
Noether gauge embedment methods. The later has led us to a
non-unitary theory while the former method has furnished the model
(\ref{sdual1}) which is dual to the Maxwell-Proca theory in
arbitrary $D$ dimensions without however, manifest $U(1)$ gauge
symmetry.

In section 3 we have applied another procedure. After starting with a rather general second-order (in
derivatives) action, see (\ref{ansatz}), involving quadratic terms in the vector and tensor fields, we have
integrated in the path integral over the tensor field and obtained an effective action for the vector field.
Requiring that the effective vector theory be $U(1)$ invariant and contain only one massive physical particle in
the spectrum we have deduced a set of constraints for the couplings. In particular, we have derived the  $U(1)$
invariant unitary models (\ref{sg}) and (\ref{sdual2}). However, after a local redefinition of the vector field
$A_{\mu} \to A_{\mu} - \p_{\mu}W/D $, involving the trace $W=W^{\mu}_{\,\mu}$, the manifest $U(1)$ symmetry is
lost very much like in the usual Stueckelberg formalism although our action is rather different than the usual
\stuck form of the Maxwell-Proca theory. In our case the trace $W$  is eaten up by the vector field which
becomes massive. We have also tried other solutions of the constraint equations but it turns out that it is
always possible to eat up the trace and end up without explicit $U(1)$ symmetry. We have given a symmetry
argument explaining that point. Clearly, one might try to include higher derivative (above second-order) terms
in the action but they are expected to jeopardize unitarity.

Regarding the TMBF model, the key difference seems to be that the $U(1)$ gauge symmetry of the vector field does
not need to be compensated by any transformation of the auxiliary two-form field unlike the case investigated
here where the symmetric rank-2 tensor must transform nontrivially.

 We are currently investigating a non-abelian extension of our
 results. Moreover, in \cite{pr2} the coupling of
  higher spin particles to the the electromagnetic field has been
  studied leading to some apparently universal conclusions. In \cite{pr2} the usual \stuck formalism has been employed.
  It is desirable to check the universality of their results via an
  alternative gauge invariant formulation for massive
  particles as given here. We are working on a generalization of
  our approach to higher spin charged particles

\section{Acknowledgements}

 We thank Alvaro de S.
Dutra for discussions and for drawing our attention to reference \cite{an}. The work of D.D. is partially
supported by CNPq while E.L.M is supported by CAPES. E.L.M. thanks the ICTP at Trieste-Italy for hospitality
where part of this work was carried out.

\end{document}